\documentclass[12pt,a4paper,titlepage,oneside]{article}
\usepackage{cite, rotate}
\usepackage{amsmath}
\usepackage[dvips]{graphics}
\usepackage[dvips]{graphicx}
\usepackage[bf]{subfigure}
\begin{document}


\begin{center}
\textbf{\Large{Grand Antiprism and Quaternions}}
\end{center}

{\bf Mehmet Koca$^1$, Mudhahir Al-Ajmi$^1$, Nazife Ozdes Koca$^1$} \\
$^1$Department of Physics, College of Science, Sultan Qaboos  \\
University, P. O. Box 36, Al-Khoud, 123 Muscat, Sultanate of Oman \\

{\bf Email}: kocam@squ.edu.om, mudhahir@squ.edu.om, nazife@squ.edu.om

\section*{Abstract}
  \noindent Vertices of the 4-dimensional semi-regular polytope, the
  \textit{grand antiprism} and its symmetry group of order 400 are
  represented in terms of quaternions with unit norm. It follows from
  the icosian representation of the \textbf{$E_{8} $} root system
  which decomposes into two copies of the root system of $H_{4} $. The
  symmetry of the \textit{grand antiprism} is a maximal subgroup of
  the Coxeter group $W(H_{4} )$. It is the group $Aut(H_{2} \oplus
  H'_{2} )$ which is constructed in terms of 20 quaternionic roots of
  the Coxeter diagram $H_{2} \oplus H'_{2}$. The root system of $H_{4}
  $ represented by the binary icosahedral group \textit{I }of order
  120, constitutes the regular 4D polytope 600-cell. When its 20
  quaternionic vertices corresponding to the roots of the diagram
  $H_{2} \oplus H'_{2}$ are removed from the vertices of the 600-cell
  the remaining 100 quaternions constitute the vertices of the\textit{
    grand antiprism}. We give a detailed analysis of the construction
  of the cells of the\textit{ grand antiprism} in terms of
  quaternions. The dual polytope of the \textit{grand antiprism} has
  been also constructed.

\section{Introduction}

\noindent Six regular 4D polytopes and their semi-regular Archimedean
polytopes are very interesting because their symmetries
$W(A_{4}),W(B_{4} ),W(F_{4} )$ and $ W(H_{4} )$ can be represented
with the use of finite subgroups of quaternions. There is a close
correspondence between the finite subgroups of quaternions and the
symmetries of the above polytopes [1]. Some of them occur in high
energy physics in model building either as a gauge symmetry like
$SU(5)$~[2] or as a little group $SO(9)$~[3] of the M-theory.  The
Coxeter group $W(H_{4} )$ arises as the symmetry group of the polytope
600-cell, $\{ 3,3,5\}$~[4], vertices of which can be represented by
120 quaternions of the binary icosahedral group [5, 6]. The dual
polytope 120-cell, $\{ 5,3,3\} $ with 600 vertices, can be constructed
from 600-cell in terms of quaternions. Another observation is that
they are all nested in the root system of $E_{8}$~represented by
icosians~[7, 8].

\noindent The\textit{ grand antiprism} was first constructed in 1965
by Conway and Guy [9] with a computer analysis. An antiprism is
defined to be a polyhedron composed of two parallel copies of some
particular $n-$sided polygon,connected by an alternating band of
triangles. In this paper we study the symmetry group of the\textit{
  grand antiprism} and construct its 100 vertices in terms of
quaternions. We explicitly show how these vertices form 20 pentagonal
antiprisms and 300 tetrahedra.  We organize the paper as follows. In
Section 2 we briefly discuss the correspondence between the root
system of $H_{4} $ and the quaternionic representation of the binary
icosahedral group \textit{I} and decompose it as 120=20+100 where
first 20 quaternions represent the root system of the Coxeter diagram
$H_{2} \oplus H'_{2} $ and the remaining quaternions are the vertices
of the \textit{grand antiprism}. The symmetry group of the\textit{
  grand antiprism} turns out to be the group $Aut(H_{2} \oplus H'_{2}
)$ of order 400 which can be constructed directly from the
quaternionic roots of the Coxeter diagram $H_{2} \oplus H'_{2}$.  A
simple technique has been employed in Section 3 for the construction
of the cell structures of the \textit{grand antiprism}. A projection
of the grand antiprism into three dimensional space is discussed in
Section 4. The dual polytope of the \textit{grand antiprism} has been
constructed in Section 5 where 320 vertices decompose as
320=20+200+100 under the group $Aut(H_{2} \oplus H'_{2} )$ where 20
vertices are scaled copies of the root system of the Coxeter diagram
$H_{2} \oplus H'_{2} $ lying on a sphere $S^{3} $ with radius
$\frac{1+\sqrt{5} }{2\sqrt{2} } \approx 1.14$ and 300 vertices lie on
an inner sphere $S^{3} $ with a radius of unit length. Conclusion is
given in Section 6 and an Appendix lists the decompositions of the
vertices of the Archimedian $W(H_{4} )$ solids in terms of the orbits
of $Aut(H_{2} \oplus H'_{2} )$.

\noindent 

\noindent 

\section{Construction of the root system of $H_{2} \oplus H'_{2}
$ as the maximal subset of the root system of $H_{4}$ and the
  vertices of the \textit{grand antiprism}}

\noindent \textbf{\textit{}}

\noindent Let $q=q_{0} +q_{i} e_{i} $, ($i=1,2,3)$ be a real
quaternion with its conjugate defined by $\bar{q}=q_{0} -q_{i} e_{i} $
where the quaternionic imaginary units satisfy the relations

\begin{equation} \label{GrindEQ__1_} 
e_{i} e_{j} =-\delta _{ij} +\varepsilon _{ijk} e_{k} , (i,j,k=1,2,3).    
\end{equation} 

Here $\delta _{ij} $and $\varepsilon _{ijk}$ are the Kronecker and
Levi-Civita symbols respectively and summation over the repeated
indices is implicit. Quaternions generate the four dimensional
Euclidean space where the quaternionic scalar product is defined by

\begin{equation} \label{GrindEQ__2_} 
(p,q)=\frac{1}{2} (\bar{p}q+\bar{q}p)=\frac{1}{2} (p\bar{q}+q\bar{p}).
\end{equation}

The imaginary quaternionic units $e_{i}$ can be related to the Pauli
matrices $\sigma _{i} $ by $e_{i} =-i\sigma _{i}$ and the unit
quaternion is represented by $2\times 2$ unit matrix. This
correspondence proves that the group of quaternions is isomorphic to
$SU(2)$ which is a double cover of the proper rotation group $SO(3)$.
The root system of the Coxeter diagram $H_{4}$ can be represented by
the quaternionic elements of the binary icosahedral group \textit{I
}[10] which are given as the set of conjugacy classes in Table 1. The
sets of conjugacy classes are indeed the orbits of the Coxeter group

\begin{equation} \label{GrindEQ__3_} 
W(H_{3} )=\{ [p,\bar{p}]\oplus [p,\bar{p}]^{*} ,{\; \; p}\in {I\} }
\end{equation} 

where we use the notation [8]

\begin{equation} \label{GrindEQ__4_} 
[a,b]:q\to aqb{\rm \; and\;} [a,b]^{*} 
{:q}\to {a}\bar{{q}}{b}.
\end{equation}

\begin{table}[h]
  \caption[l]{\normalsize{Conjugacy classes of the binary 
      icosahedral group $I$ represented by quaternions for the 
      cases of~1 being fixed. (Sets of conjugacy classes of $N$ elements
      fixing the elements $q$ are denoted as $N(q)$ where additional
      subscripts are used to distinguish different sets of the same type.)}}

\begin{tabular}{lll}
  \hline
  Orders of elements    &&   The sets $N(1)$ of the conjugacy classes  \\
                        &&   denoted with $N$ elements \\

   \hline
  1  & $1$ &   \\  
  2  & $-1$ &   \\  
  10 &  $12(1)_+:$ & $\frac{1}{2} (\tau \pm e_{1} \pm \sigma e_{3} ),\frac{1}{2} (\tau \pm e_{2} \pm \sigma e_{1} ),$\\
     &  &    $\frac{1}{2} (\tau \pm e_{3} \pm \sigma e_{2})$ \\  
  5  &   $12(1)_-:$ & $\frac{1}{2} (-\tau \pm e_{1} \pm \sigma e_{3} ),\frac{1}{2} (-\tau \pm e_{2} \pm \sigma e_{1} ),$  \\ 
     & &  $\frac{1}{2} (-\tau \pm e_{3} \pm \sigma e_{2})$ \\  
  10 &   $12(1)_+^\prime:$ & $\frac{1}{2} (\sigma \pm e_{1} \pm \tau e_{2} ),\frac{1}{2} (\sigma \pm e_{2} \pm \tau e_{3} ),$  \\
     &  &   $\frac{1}{2} (\sigma \pm e_{3} \pm \tau e_{1})$ \\  
  5  &   $12(1)_-^\prime:$ & $\frac{1}{2} (-\sigma \pm e_{1} \pm \tau e_{2} ),\frac{1}{2} (-\sigma \pm e_{2} \pm \tau e_{3} ),$  \\
     &  &   $\frac{1}{2} (-\sigma \pm e_{3} \pm \tau e_{1})$  \\  
  6  &   $20(1)_+:$ & $\frac{1}{2} (1\pm e_{1} \pm e_{2} \pm e_{3} ),\frac{1}{2} (1\pm \tau e_{1} \pm \sigma e_{2} ),$  \\  

     &   & $\frac{1}{2} (1\pm \tau e_{2} \pm \sigma e_{3} ),\frac{1}{2} (1\pm \tau e_{3} \pm \sigma e_{1} )$ \\  
  3  & $20(1)_-:$ & $\frac{1}{2} (-1\pm e_{1} \pm e_{2} \pm e_{3} ),\frac{1}{2} (-1\pm \tau e_{1} \pm \sigma e_{2} ),$  \\ 
     &  & $\frac{1}{2} (-1\pm \tau e_{2} \pm \sigma e_{3} ),\frac{1}{2} (-1\pm \tau e_{3} \pm \sigma e_{1} )$ \\  
  4  &  $30(1):$ & $\pm e_{1} ,\pm e_{2} ,\pm e_{3} ,\frac{1}{2} (\pm \sigma e_{1} \pm \tau e_{2} \pm e_{3} ),$ \\ 
     &  & $\frac{1}{2} (\pm \sigma e_{2} \pm \tau e_{3} \pm e_{1} ),\frac{1}{2} (\pm \sigma e_{3} \pm \tau e_{1} \pm e_{2} )$ \\  
  
  \hline
\end{tabular}
\end{table}

\noindent In Table 1 $\tau =\frac{1+\sqrt{5} }{2} ,{\rm \; \; \; \; \;
}\sigma {\rm =}\frac{1-\sqrt{5} }{2} $ and they satisfy the relations
$\tau ^{2} =\tau +1,{\rm \; \; }\sigma ^{{\rm 2}} =\sigma +1,{\rm \;
  \; }\tau {\rm +}\sigma {\rm =1,\; \; }\tau \sigma {\rm =-1.}$ The
notation \# $(1)$ denote the set of elements of the
conjugacy classes which also represent the orbits of the Coxeter
group $W(H_{3} )$ fixing the element~1.

\noindent Let $b=\frac{1}{2} (\tau +\sigma e_{1} +e_{2} )\in
12(1)_{+}$ and $b^5 =-1$ represent the two simple roots of the Coxeter
group $H_{2} $ with unit norm. Then the quaternions $e_{3}
b=\frac{1}{2} (-e_{1} +\sigma e_{2} +\tau e_{3} ){\rm \; and\; \;
}e_{3} {\rm b}^{{\rm 5}} =-e_{3}$ represent the simple roots of the
Coxeter diagram $H'_{2}$ orthogonal to the roots of $H_2$. The Coxeter
diagram of $H_{2} \oplus H'_{2}$ with its simple roots are depicted in
Figure~1.

\begin{figure}[b]
\begin{center}
  \includegraphics{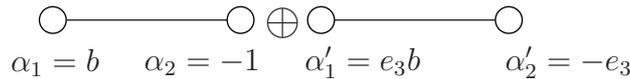}
\end{center}
\caption{Coxeter diagram of $H_{2} \oplus H'_{2}$ with quaternionic simple roots.}
\end{figure}

\noindent The reflection generators at these simple roots are given by [11]

\begin{equation} \label{GrindEQ__5_} 
[b,-b]^{*} ,{\rm \; \; \; \; \; [1,-1]}^{*} ,{\rm \; \; \; }[e_{3} b,-e_{3} b]^{*} ,{\rm \; \; \; \; \; [}e_{3} {\rm ,-}e_{3} {\rm ]}^{*} .
\end{equation}

These generators will lead to the following 20 quaternionic roots of
the Coxeter diagram $H_{2} \oplus H'_{2}$

\noindent 

\begin{equation} \label{GrindEQ__6_} 
H_{2} \oplus H'_{2} =\{ b^{m} ,{\rm \; \; e}_{{\rm 3}} b^{m}\},~(m=0,1,...,9).                                               
\end{equation}

\noindent This set of quaternions form a maximal subgroup, the
dicyclic group of order 20, in the binary icosahedral group \textit{I}
[4]. The generators in~(\ref{GrindEQ__5_}) generate the group $W(H_{2}
)\times W(H'_{2} )$, each of which, is a dihedral group of order~10.
It is more appropriate to represent each dihedral group with one
rotation generator and one reflection generator which can be written
as

\begin{equation} \label{GrindEQ__7_} 
W(H_{2} )=\{ [b,b],[1,-1]^{*} \} ;{\rm \; \; \; \; \; \; \; \; \; \; }W(H'_{2} )=\{ [\bar{b},b],[e_{3} ,-e_{3} ]^{*} \}.      
\end{equation}

\noindent The direct product group is of order 100. One may simply
note that the generator $[e_{3} ,1]$ or $[1,e_{3}]$ permutes the
simple roots of $H_{2} \oplus H'_{2}$. Actually these generators leave
the Cartan matrix of the diagram $H_{2} \oplus H'_{2} $ invariant.
Without loss of generality we take the generator $[e_{3} ,1]$ as the
new generator which generates the cyclic group $C_{4}$ of order~4. The
group $C_{4}$ is the normalizer of the group $W(H_{2} )\times
W(H'_{2} )$ in the Coxeter group $W(H_{4} )$. An extension of the
group $W(H_{2} )\times W(H'_{2} )$ by the group $C_{4}$ is a
semi-direct product of the two groups which is the group

\noindent 

\begin{equation} \label{GrindEQ__8_} 
Aut(H_{2} \oplus H'_{2} )\approx \{ W(H_{2} )\times W(H'_{2} )\} :C_{4}      
\end{equation}

of order 400 and is one of the five maximal subgroups of the Coxeter
group $W(H_{4})$~[11].

\noindent The group elements now can be written more concisely as

\begin{equation} \label{GrindEQ__9_} 
Aut(H_{2} \oplus H'_{2} )=\{ [p,q]\oplus [p,q]^{*} ;{\; \; p,q}\in {\rm (}H_{2} \oplus H'_{2} )\}  .                 
\end{equation} 

\noindent It is represented by the pairs of quaternions taking values
from the root system of the Coxeter diagram $H_{2} \oplus H'_{2}$.
This is the quaternionic representation of the ionic diminished
Coxeter group $[10, 2^+,10]$~[6].

\noindent 

\section{Construction of vertices of the \textit{grand antiprism}}

\noindent 

\noindent The set of roots of the Coxeter graph $H_{2} \oplus H'_{2}$
represents the vertices of two decagons in two orthogonal planes
sharing only the origin as the common point. For a better
clarification let us define a new set of quaternionic units

\begin{equation} \label{GrindEQ__10_} 
1'=1,{\rm \; \; \; \; \; \; }{e}'_{{\rm 1}} =\frac{\sigma e_{1} +e_{2} }{\sqrt{2+\sigma } } ,{\rm \; \; \; \; }{e}'_{{\rm 2}} =e_{3} {e}'_{{\rm 1}} =\frac{-e_{1} +\sigma e_{2} }{\sqrt{2+\sigma } } {\rm ,\; \; }{e}'_{{\rm 3}} =e_{3}  .             
\end{equation} 
Then the set of roots in (\ref{GrindEQ__6_}) can be written as 

\begin{equation} \label{GrindEQ__11_} 
\{ \exp {(\frac{\pi }{5} e'_{1} )^{m} } ,{\rm \; \; \; \; \;} {e}_{{\rm 3}} \exp {(\frac{\pi }{5} e'_{1} )^{m} } \} ,{\; \; (m=0,1,...,9)}. 
\end{equation}

\noindent This indicates that the roots of the first $H_{2} $ and the
origin define a plane, say plane \textbf{A}, generated by the unit
vectors 1, $e'_{1}$ and the second set of roots with the origin define
the plane \textbf{B}, generated by the unit vectors $e'_{2} {\rm \;
  and\;} \; e_{\rm 3}$. The group element $[b,b]$ acts as a rotation
of order 5 in the plane \textbf{A} and the element $[\bar{b},b]$ acts
as a rotation of order 5 in the plane \textbf{B}.  In what follows we
prove that the remaining set of 100 quaternions obtained by deleting
those quaternions in (\ref{GrindEQ__6_}) from the set \textit{I}, can
be written compactly as vertices of the grand antiprism (GA):

\begin{equation} \label{GrindEQ__12_}
  {\rm \; GA=\{}{b}^{{m}} cb^{n} ,{b}^{{m}} e_{3} cb^{n} \}
  ,{\; \; (m,n=0,1,...,9)}.
\end{equation}
          
\noindent Here $c=\frac{1}{2} (\tau -\sigma e_{1} +e_{2} )\in
12(1)_{+} $ is one of those nearest quaternions to $b$ in \textit{I}.
In reference [10] we showed that the sets of quaternions in Table 1
represent the vertices of some polyhedra possessing the icosahedral
symmetry $W(H_{3} )$. Each set of 12 elements represents an
icosahedron, each set of 20 elements represents a dodecahedron and the
set with 30 elements represents an icosidodecahedron. The unit
quaternions $\pm 1$ represent the poles of the sphere $S^{3}$. It is
clear that these polyhedra have the vertices at the intersections of
the parallel hyperplanes orthogonal to the unit quaternion 1 with the
sphere $S^{3}$. For example, the set $12_{+}(1)$ can be plotted in a
space where the imaginary unit quaternions represent the ordinary $x,~
y,~z$ axes. This means we are in the space represented by the
intersection of the sphere $S^{3}$ with the hyperplane orthogonal to
the unit quaternion 1 and at a distance $\frac{\tau }{2}$ from the
origin.

\noindent In order to understand the 600-cell, the set $12(1)_{+}$
plays the most important role. The quaternions in the set $12(1)_{+} $
constitute the vertices of an icosahedron with the shortest edge
length $-\sigma $.  They can be decomposed into 20 sets of triads [10]
where each triad represents an equilateral triangle. The edge length
between the unit quaternion 1 and any one of those quaternions in the
set $12(1)_{+} $ is also $-\sigma$ so that the quaternion 1 and the
quaternions of the set $12(1)_{+}$ constitute 20 tetrahedra with a
common vertex represented by 1. Table~1 is just one decomposition out
of 120 different decompositions where any quaternion $q\in I$ can be
chosen instead of~1. This is because the group $W(H_{3} )$ can be
embedded in 120 different ways in the group $W(H_{4} )$. In each case
the group $W(H_{3} )$ leaves one quaternion, \textit{q} say, invariant
which can be represented by the conjugate group $W(H_{3})^{q} =\{
[I,\bar{q}\bar{I}q]\oplus [I,q \bar{I} q]^{*} \}$~[10].  The 20
tetrahedra having the quaternion \textit{q} as a common vertex would
be represented by the set of quaternions $q{\rm \; \; and\;
  q(12(1)}_{{\rm +}} )={\rm (12(1)}_{{\rm +}} )q=12(q)_{+}$. This
proves that the 600-cell consists of $\frac{20\times 120}{4} =600$
tetrahedra.

\noindent The set $12(1)_{+} $ can be written in terms of the
quaternions $b{\rm \; and\; c}$ as follows:

\noindent 

\begin{equation} \label{GrindEQ__13_} 
\begin{array}{l} {{ \; \; \; \; \; \; \; \; \; \; \; \; \; \; \; \; \; \; \; \; \; \; \; \; \; b}} \\ {{ \; \; \; \; \; \; \; \; c\; \; \; \; \; \; \; \; \; }\bar{{ b}}{ cb\; \; \; \; \; \; \; }\bar{{ b}}^{{ 2}} cb^{2} { \; \; \; \; \; b}^{{ 2}} c\bar{b}^{2} { \; \; \; \; \; bc}\bar{{ b}}} \\ {} \\ {{ c}\bar{{ b}}{\; \; \; \; \; \; \; \; \; }\bar{{ b}}{ c\; \; \; \; \; \; \; \; \; }\bar{{ b}}^{{ 2}} cb{ \; \; \; \; \; \; \; -b}^{{ 2}} cb^{2} { \; \; \; \; \; \; bc}\bar{{ b}}^{{ 2\; }} { \; \; \; \; \; \; }} \\ {{ \; \; \; \; \; \; \; \; \; \; \; \; \; \; \; \; \; \; \; \; \; \; \; \; \; }\bar{{ b}}} \end{array} 
\end{equation}

\noindent Indeed the third line is the set of five quaternions
$(\bar{b}^{m} \bar{c}b^{m},~m=\{0,1,2,3,4\})$ obtained by rotating the
quaternion $\bar{c}$ around the quaternion \textit{b.  } This
assertion can be checked through the relations

\noindent 

\begin{equation} \label{GrindEQ__14_} 
b^{2} cb^{2} =-\bar{c},{\; \; \; \; }\bar{{ b}}{ cb=cb}\bar{{ c}}{ ,\; \; \; }\bar{{ c}}{ bc=bc}\bar{{ b}}{ ,\; \; \; \; e}_{{ 3}} b=\bar{b}e_{3} ,{ \; \; \; e}_{{ 3}} c=\bar{c}e_{3}  .         
\end{equation}

\noindent As we stated earlier they are the vertices of an icosahedron
when plotted in the ordinary 3D space represented by $x,~y,~z$ as
depicted in Figure~2.

\begin{figure}[h]
\begin{center}
  \includegraphics[height=4cm]{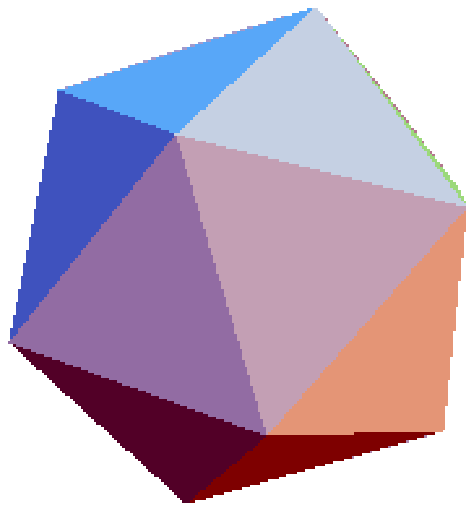}
\end{center}
\caption{The icosahedron of the quaternions of Eq.\ref{GrindEQ__13_}}
\end{figure}

\noindent If one imagines that the quaternions $b$ and $\bar{b}$
represent the top and the bottom vertices of the icosahedron in
(\ref{GrindEQ__13_}) then those quaternions in the second line
constitute the vertices of a pentagon nearest to the quaternion $b$
and those in the third line represent the bottom pentagonal plane
nearest to the quaternion $\bar{b}$. One can easily show that the
quaternions representing the edges of the pentagons in both top and
bottom planes can be written as follows

\noindent 
\begin{equation} \label{GrindEQ__15_} 
-\sigma (\alpha e'_{2} +\beta e_{3} ),{\rm \; \; }\alpha ^{{\rm 2}} +\beta ^{2} =1 .                                         
\end{equation}

Indeed these two planes are parallel to the plane \textbf{B } which is
orthogonal to the plane~\textbf{A} generated by 1 and $e'_{1}$.  Since
$b$ and $\bar{b}$ are removed from the set of quaternions in
(\ref{GrindEQ__13_}) what remains is the pentagonal antiprism with 10
vertices which is depicted in Figure~3.

\begin{figure}[h]
\begin{center}
  \includegraphics[height=3.5cm]{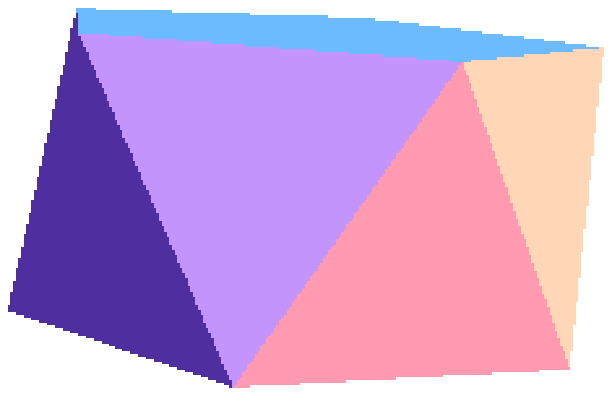}
\end{center}
\caption{Pentagonal antiprism represented by the quaternionic vertices
  $(\bar{b}^{m} cb^{m} ,{\rm \; \; }\bar{b}^{m} \bar{c}b^{m} {\rm ),\;
    \;} {(m=0,1,2,3,4).\; \; \; }$}
\end{figure}

\noindent There is now a simple mechanism to generate 10 pentagonal
antiprisms which form a ring. Before proceeding further we note that
all quaternions on the second line of (\ref{GrindEQ__13_}) are
obtained from the vertex \textit{c} by a rotation in the plane
\textbf{B} by an angle $72^{0}$. Similarly the quaternions on the
third line of (\ref{GrindEQ__13_}) are obtained by a $72^{0}$ rotation
of $\bar{c}$ in the plane parallel to the plane \textbf{B}.

\noindent This shows that the edge length of the pentagon is
$-\sigma$. \\
Now we discuss how to obtain those 50 quaternions constituting the
first ring of 10 pentagonal anti-prisms. Let us consider the diagram
in Figure~4

\begin{figure}[h]
\begin{center}
  \includegraphics[height=3.5cm]{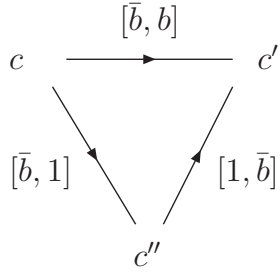}
\end{center}
\caption{Diagrammatic representation of some of the group elements of $Aut(H_{2} \oplus H'_{2} )$}
\end{figure}

\noindent This shows that the group element $[\bar{b},b]$ rotates the
quaternions by an angle $72^{0}$ in the plane \textbf{B}, in other
words, in the ten planes parallel to \textbf{B}. The group element
$[\bar{b},1]$ transforms a given quaternion in the `upper plane' to a
quaternion in the `lower plane' which is already rotated by $36^{0}$
degrees with respect to the `upper plane'. The group element $[1,b]$
lifts one element from the bottom plane to the upper plane after a
$36^{0}$ rotation. It is interesting to note that just as the sum of
these vectors $(c''-c)+(c'-c'')=c'-c$, the group elements satisfy a
similar relation, namely, $[\bar{b},1][1,b]=[\bar{b},b]$. Inverses of
these group elements reverse the directions of the arrows. The
quaternions $c,{\rm \; \; \; }{c}'{\rm =}\bar{{b}}{cb,\; \; \;
}{c}''{\rm =}\bar{{b}}{c}$ form an equilateral triangle. Applying
$[1,b]$ ten times on any quaternion will bring it back to the original
quaternion so that they define a decagon.  Similarly the action of
$[\bar{b},1]$ on any quaternion ten times will generate another plane
of decagon. It is clear that we obtain the following ring of ten
pentagonal antiprisms provided the quaternions $b^{m},~(m=0,1,...,9)$
are removed from the set \textit{I}.

\begin{figure}[h]
\begin{center}
  \includegraphics[height=8cm]{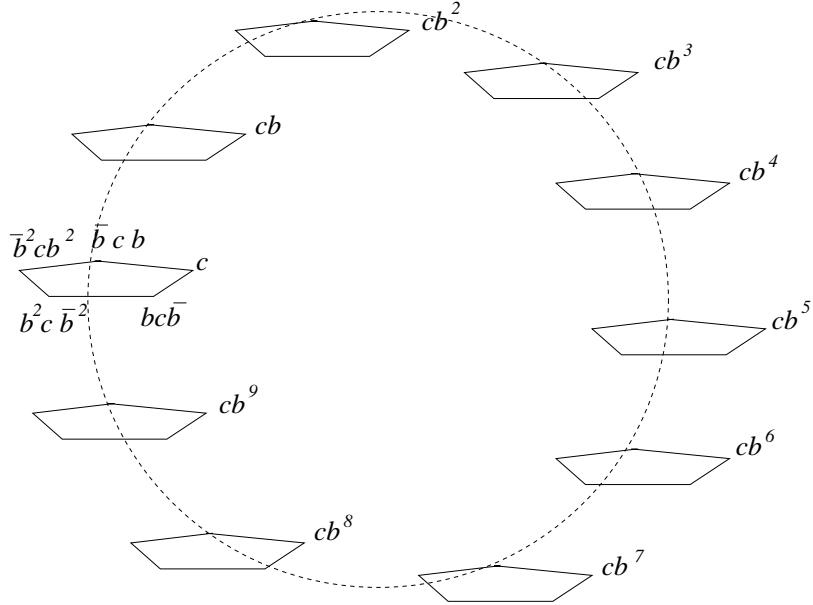}
\end{center}
\caption{One ring of pentagonal antiprism. The quaternions at the
  vertices of the pentagons are obtained by $\frac{2\pi}{5}$ rotations
  around the vertex $b$.}
\end{figure}

Figure~5 indicates that when starting with the quaternion \textit{c}
we obtain the set of 50 quaternions

\begin{equation} \label{GrindEQ__16_} 
R_{1} =\{ b^{m} cb^{n} {\rm \; \} ,\; \; \; \; \;} {(m,n=0,1,...,9)}  .                             
\end{equation} 

\noindent It is clear from Figure~5 that the vertices of two pentagonal
antiprisms in which the quaternion \textit{c} is a common vertex can
be written as follows:

\noindent 

\begin{equation} \label{GrindEQ__17_}
\begin{array}{l} {{ \; \; \; \; \; \; \; \; \; \; }} \\ {{ \; \; \; \; \; \; \; \; \; \; \; \; cb,\; \; \; \; \; \; }\bar{{ b}}{ cb}^{{ 2}} ,{ \; \; \; -}\bar{{ b}}^{{ 2}} { c}\bar{{ b}}^{{ 2}} { ,\; \; \; \; \; b}^{{ 2}} { c}\bar{{ b}}{ ,\; \; \; \; \; \; bc\; \; \; \; \; \; \; }} \\ {{ \; \; \; \; \; \; \; \; c,\; \; \; \; \; \; }\bar{{ b}}{ cb,\; \; \; \; \; }\bar{{ b}}^{{ 2}} { cb}^{{ 2}} { ,\; \; \; \; \; b}^{{ 2}} { c}\bar{{ b}}^{{ 2}} { ,\; \; \; \; bc}\bar{{ b}}} \\ {{ \; \; \; \; }\bar{{ b}}{ c,\; \; \; \; \; }\bar{{ b}}^{{ 2}} { cb,\; \; \; -b}^{{ 2}} { cb}^{{ 2}} { ,\; \; \; bc}\bar{{ b}}^{{ 2}} { ,\; \; \; \; c}\bar{{ b}}{ \; \; \; \; }} \end{array}                                 
\end{equation}

\noindent Note that the quaternions at the upper two lines can be
obtained from those at the lower two lines by a right multiplication
by the quaternion \textit{b}. In other words, the set of quaternions
in the upper two lines are the elements of the set $12_{+} (b)$ from
which $1$ and $b^{2} $ are deleted. They constitute the pentagonal
antiprism in the space of intersection of the hyperplane orthogonal to
the quaternion \textit{b} and the sphere $S^{3}$. We note that the
quaternions at the bottom line are the quaternionic conjugates of the
quaternions at the middle line. Therefore the center of the lower
pentagonal antiprism is at the point $\frac{\tau }{2} 1$ and the center
of the upper pentagonal antiprism is at the quaternion $\frac{\tau }{2}
b$. If we keep multiplying the quaternions representing a given
pentagonal antiprism by \textit{b} we create other quaternionic
vertices. The next layer of vertices are the set of $12_{+} (b^{2} )$
with the omission of $b$ and $b^{3}$ where the center is represented by
$\frac{\tau }{2} b^{2} $. We have altogether ten pentagonal antiprisms
in this ring represented by the sets $12_{+} (b^{m} ),{\rm \; \;
}$ where the vertices $b^{m} ,{m=0,1,...,9}$ are deleted from
the sets and the centers of the pentagonal antiprisms are represented
by ten quaternions $\frac{\tau }{2} b^{m}$ . We note that the action
of the element $[1,b]$ corresponds to the rotation of the hyperplane
whose intersection with the sphere $S^{3}$ is a pentagonal antiprism.

\noindent When we act by the group element $[e_{3} ,1]$ on the set of
quaternions in the diagram of Figure~4 and find the conjugate elements
of the group elements we obtain the diagram in Figure~6.

\begin{figure}[h]
\begin{center}
 \includegraphics[height=3.5cm]{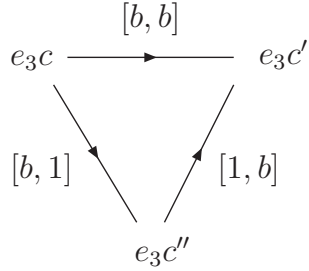}
\end{center}
\caption{Diagrammatic representation of the action of some group
  elements on the set of vertices of the second ring of the pentagonal
  anti-prisms.}
\end{figure}

\noindent A similar consideration will lead to the set of quaternions

\begin{equation} \label{GrindEQ__18_} 
R_{2} =\{ b^{m} e_{3} cb^{n} \} ,{\rm \; \; \; \; \; } {(m,n=0,1,...,9)} 
\end{equation} 
which represents the second ring of vertices of ten pentagonal
antiprisms similar to those depicted in Figure~5. The two sets of
rings satisfy the relations

\noindent 

\begin{equation} \label{GrindEQ__19_} 
e_{3} R_{1} =R_{1} e_{3} =R_{2} ,{\rm \; \; }e_{3} R_{2} =R_{2} e_{3} =R_{1}.                              
\end{equation}

\noindent It is clear from these discussions that the centers of the
pentagonal antiprisms in both rings are represented by the set of 20
quaternions

\noindent 

\begin{equation} \label{GrindEQ__20_} 
\frac{\tau }{2} b^{m} ,{\rm \; \; \; \; }\frac{\tau }{2} e_{3} b^{m} {\rm ,\; \; \;} {(m=0,1,...,9)\; }.                                         
\end{equation}

\noindent 

\noindent The \textit{grand antiprism} has 300 tetrahedral cells,
details of which we discuss below. In order to understand the cell
structure of a polytope one should check first the cell structures
surrounding a given vertex, say, vertex \textit{c}. We have already
seen that the vertex \textit{c} is surrounded by two pentagonal
antiprisms illustrated in Eq.~(\ref{GrindEQ__17_}). To understand the
nature of the tetrahedra linked to the vertex \textit{c} we look at
the set of quaternions $12_{+} (c)=12_{+} (1)c=c(12_{+}(1))$ which can
be written as follows:

\begin{equation} \label{GrindEQ__21_} 
\begin{array}{l} {{ \; \; \; \; \; \; \; \; \; \; \; \; \; \; \; \; \; \; \; \; \; \; \; \; \; \; \; \; \; \; \; cb}} \\ {{ \; \; \; \; \; \; c}^{{ 2}} { ,\; \; \; c}\bar{{ b}}{ cb,\; \; \; \; c}\bar{{ b}}^{{ 2}} { cb}^{{ 2}} { ,\; \; \; b,\; \; \; \; \; \; cbc}\bar{{ b}}} \\ {} \\ {{ \; \; \; \; \; \; \; \; c}\bar{{ b}}{ c,\; \; \; \; \; c}\bar{{ b}}^{{ 2}} { cb,\; \; \; \; \; \; \; 1,\; \; \; \; cbc}\bar{{ b}}^{{ 2}} { ,\; \; \; \; \; \; c}^{{ 2}} \bar{{ b}}{ \; \; }} \\ {{ \; \; \; \; \; \; \; \; \; \; \; \; \; \; \; \; \; \; \; \; \; \; \; \; \; \; \; \; \; \; \; \; \; \; c}\bar{{ b}}} \\ {{ \; }} \end{array}    . \end{equation}

\noindent We note that we should remove two quaternions 1 and $b$ from
the set of~(\ref{GrindEQ__21_}) since they are the roots of $H_{2} $.
It is easy to see that the remaining 10 vertices form 12 equilateral
triangles. When they are connected to the vertex $c$ we obtain 12
tetrahedra. This proves that the vertex $c$ is surrounded by two
pentagonal antiprisms and 12 tetrahedra. Let us write those 10
quaternions explicitly

\begin{equation} \label{GrindEQ__22_} 
\begin{array}{ll} 
q_{1} =\frac{1}{2} (1+e_{1} +e_{2} -e_{3} ), & q_{2} =\frac{1}{2} (1+\tau e_{2} -\sigma e_{3}), \\ 
q_{3} =\frac{1}{2} (\tau +e_{1} -\sigma e_{3}), & q_{4} =\frac{1}{2} (1+\tau e_{1} -\sigma e_{2} ), \\
q_{5} =\frac{1}{2} (\tau -\sigma e_{2} +e_{3}), & q_{6} =\frac{1}{2} (-\sigma +e_{1} +\tau e_{2}), \\
q_{7} =\frac{1}{2} (\tau +e_{1} +\sigma e_{3}), & q_{8} =\frac{1}{2} (\tau -\sigma e_{2} -e_{3} ), \\
 q_{9} =\frac{1}{2} (1+e_{1} +e_{2} +e_{3}), & q_{10} =\frac{1}{2} (1+\tau e_{2} +\sigma e_{3})
\end{array} 
\end{equation} 

\noindent They belong to the following sets of rings of pentagonal
antiprisms

\begin{equation} \label{GrindEQ__23_} 
(c,{ \; q}_{{2}} ,{ \; q}_{{3}} ,{ \; q}_{{5}} ,{ \; q}_{{7}} ,{ \; q}_{{8}} ,{ \; q}_{{10}} )\in R_{1} ;{ \; \; \; \; (q}_{{1}} ,{ \; q}_{{4}} ,{ \; q}_{{6}} ,{ \; q}_{{9}} )\in R_{2} .       
\end{equation}

\noindent The 12 tetrahedra can be classified as follows:

\noindent Four tetrahedra: 

\begin{equation} \label{GrindEQ__24_}
(c,q_{2} ;q_{6} ,q_{9} ),{ \; \; (c,\; q}_{{3}} ;q_{4} ,q_{9} ),{ \; \; (c,q}_{{7}} ;q_{1} ,q_{4} ),{ \; \; (c,q}_{{10}} ;q_{1} ,q_{9} ){ \; }
\end{equation} 

\noindent Six tetrahedra: 
\begin{equation} \label{GrindEQ__25_}
\begin{array}{l}
{ (c,q}_{{2}} ,q_{5} ;q_{9} ),{ \; }(c,q_{2} ,q_{10} ;q_{6} ),{ \; (c,q}_{{3}} ,q_{5} ;q_{9} ),{ \; (c,\; q}_{{3}} ,q_{7} ;q_{4} ),\\ { \; (c,\; q}_{{7}} ,q_{8} ;q_{1} ),{ \; \; (c,q}_{{8}} ,q_{10} ;q_{1} )
\end{array}
\end{equation} 

\noindent Two tetrahedra:
\begin{equation} \label{GrindEQ__26_} 
(c;q_{1} ,q_{4} ,q_{6} ),{ \; \; (c;q}_{{4}} ,q_{6} ,q_{9} ).
\end{equation} 

\noindent It is clear from these considerations that the first four
tetrahedra in~(\ref{GrindEQ__24_}) have vertex structures such that
two vertices are on the ring $R_{1}$ and two vertices are on the ring
$R_{2}$. The next six tetrahedra in~(\ref{GrindEQ__25_}) have vertex
structures such that three vertices forming an equilateral triangle on
the ring $R_{1}$ and one vertex on the ring $R_{2}$.  Similarly the
last two tetrahedra in~(\ref{GrindEQ__26_}) have the following
structure: one vertex on the ring $R_{1}$ and three vertices on the
ring $R_{2}$.  When we consider the tetrahedral structures formed by
the quaternion $e_{3} c$ and the set of quaternions $12_{+} (e_{3}
c)$, the above structures will be reversed because multiplication by
the quaternion $e_{3}$ reverses the sets of rings $e_{3} :{\rm \; \;
  \; R}_{{\rm 1}} \leftrightarrow R_{2}$. When we consider the 24
tetrahedra together we see that they split as 24=8+8+8. Since for one
vertex of the \textit{grand antiprism} we have 12 tetrahedra and each
tetrahedron has 4 vertices then the \textit{grand antiprism} has
$\frac{12\times 100}{4} =300$ tetrahedra which split as 300=100+200.
This shows that 100 tetrahedra have the structure such that two
vertices are on the ring $R_{1}$ and two vertices are on the ring
$R_{2} $. The 200 tetrahedra have the structure such that each of 100
tetrahedra has three vertices on the ring $R_{1}$ and the fourth
vertex on the ring $R_{2}$ and each of the next 100 tetrahedra has one
vertex on the ring $R_{1}$ and three vertices on the ring $R_{2} $.
\noindent The ten quaternions in (\ref{GrindEQ__22_}) constitute the
vertices of the vertex figure of the grand antiprism (see Figure~7) as
they are the nearest vertices to the vertex~$c$. They are on the
sphere $S^{2}$ determined as the intersection of the sphere $S^{3}$
and the hyperplane orthogonal to the quaternion~$c$. To plot the
vertices in Figure~7 in 3D we introduce a new set of orthogonal basis
of quaternionions defined by

\begin{equation} \label{GrindEQ__27_} 
\begin{array}{l} {d_{0} =c=\frac{1}{2} (\tau -\sigma e_{1} +e_{2} ),{\rm \; \; \; \; \; \; \; \; \; \;} {d}_{{\rm 1}} =e_{1} c=\frac{1}{2} (\sigma +\tau e_{1} +e_{3} ),{\rm \; }} \\ {{d}_{{\rm 2}} =e_{2} c=\frac{1}{2} (-1+\tau e_{2} +\sigma e_{3} ),{\rm \; \; \;} {d}_{{\rm 3}} =e_{3} c=\frac{1}{2} (-e_{1} -\sigma e_{2} +\tau e_{3} ).{\rm \; }} \end{array} 
\end{equation} 
\textit{}

\noindent In the new basis a quaternion $q$ would read

\noindent 

\[q=q_{0} d_{0} +q_{1} d_{1} +q_{2} d_{2} +q_{3} d_{3} \equiv (q_{0}
,q_{1} ,q_{2} ,q_{3} )\]

\noindent and the 10 quaternions above all have first components equal
to $\frac{\tau }{2}$. If their first components are deleted and an
overall factor $\frac{1}{2}$ is omitted then they can be written as

\noindent 

\begin{equation} \label{GrindEQ__28_} 
(0,\pm \sigma ,\pm 1),{\rm \; \; (-}\sigma {\rm ,}\pm {\rm 1,0),\; \; (}\sigma {\rm ,1,0),\; \; (1,0,}\pm \sigma {\rm ),\; \; (-1,0,}\sigma {\rm )}.                             
\end{equation}

\noindent These are the vertices of the solid called {\it dissected
  icosahedron} shown in Figure~7.

\begin{figure}[h]
\begin{center}
\includegraphics[height=4cm]{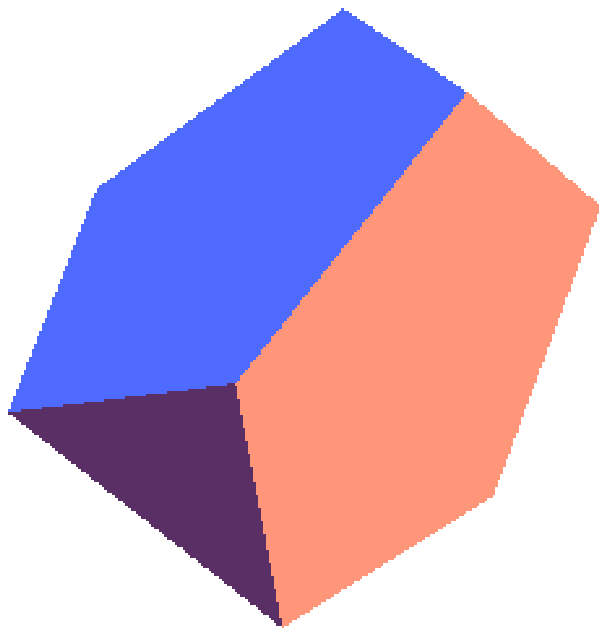}
\end{center}
\caption{Dissected icosahedron (Vertex figure of the \textit{grand
    antiprism})}
\end{figure}

\noindent The quaternions~1 and \textit{b} which were deleted can be
represented by $(-1,0,\sigma)$, $(\sigma,-1,0)$ respectively in the
new basis. With the inclusion of these two vertices to the vertices
in~(\ref{GrindEQ__28_}) one would obtain an icosahedron~[12].

\noindent Before we end this section we determine the symmetry of the
vertex figure. The subgroup of the group $Aut(H_{2} \oplus H'_{2} )$
fixing the quaternion~$c$, in other words, leaving the hyperplane
orthogonal to $c$ invariant is the symmetry of the vertex figure.
Therefore it suffices to determine the subgroup which leaves the
quaternion $c$ invariant. One can check that it is the group generated
by two commuting group elements $[e_{3} ,-e_{3} ]^{*}$ and
$[\bar{b}^{2} ,-\bar{b}^{2} ]^{*}$. So it is the Klein's four-group
$C_{2} \times C_{2}$ of order~4. The vertices of the {\it dissected
  icosahedron} which are left invariant under the action of the group
$C_{2} \times C_{2}$ are given as sets of quaternions in the brackets
as follows:

\noindent 

\begin{equation} \label{GrindEQ__29_} 
(q_{1} ,q_{9} ),{ \; (q}_{{2}} {,q}_{{3}} {,q}_{7}, {q}_{10}),\; (q_{4} ,q_{6}),(q_{5},q_{8}).         
\end{equation}

\section{Projection of the grand antiprism into 3D space}

\noindent Let us consider the intersection of the hyperplanes
orthogonal to the unit quaternion 1 with the three dimensional sphere
$S^{3}$ of unit radius consisting of 120 vertices of the 600-cell
displayed in Table~1. When the set of vectors $b^{m} ,{\rm \; e}_{{\rm
    3}} b^{m}$ ($m=0,1,2,...,9$) are removed from the set of
quaternions of Table~1, we obtain several pentagonal antiprisms some
of which are not regular. The regular pentagonal prisms follow from
the sets $12_{\pm }(1)$ and $12'_{\pm }(1)$ when $\pm {b,\; \; }\pm
\bar{{b}}{\rm ,\; \; }\pm {b}^{{\rm 2}} {\rm ,\; \; }\pm
\bar{{b}}^{{\rm 2}} {\rm \; \; \; }$ are removed from the sets.  The
vertices of four pentagonal antiprisms can be compactly written as;

\begin{equation} \label{GrindEQ__30_} 
\begin{array}{l} {(\bar{b}^{m} cb^{m} ,{\; }\bar{b}^{m} \bar{c}b^{m} {),\; \; \; }(-\bar{b}^{m} cb^{m} ,{\; -}\bar{b}^{m} \bar{c}b^{m} {),\; \; \; \; }} \\ \\ {(\bar{b}^{m} c^{2} b^{m} ,{\; }\bar{b}^{m} \bar{c}^{2} b^{m} {),\; \; \; }(-\bar{b}^{m} c^{2} b^{m} ,{\; -}\bar{b}^{m} \bar{c}^{2} b^{m} {),\; \; (m=0,1,2,3,4)\; }}. \end{array}    
\end{equation} 

\noindent They are in the intersections of the sphere $S^{3}$ with the
parallel hyperplanes orthogonal to the unit quaternion 1 and with real
coordinates at the distances $\pm \frac{\tau }{2}$ and $\pm
\frac{\sigma }{2}$. One of these pentagonal antiprism has been
depicted in Figure~3. Similarly, at distances $\pm \frac{1}{2}$ from
the origin along the quaternion 1 we have two dodecahedra, namely the
sets $20_{+}$ and $20_{-}$ one of which is depicted in Figure~8.

\begin{figure}[h]
\begin{center}
  \includegraphics[height=4cm]{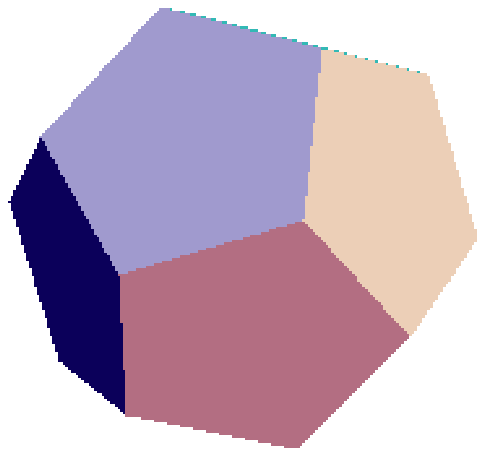}
\end{center}
\caption{Dodecahedron obtained from the projection of the
  \textit{grand antiprism}}
\end{figure}

One of the sets, say, $20_{+}$ now decomposes into two sets as $20_{+}
=10_{+} +10'_{+}$ under a subgroup, of order 20, of the group
$Aut(H_{2} \oplus H'_{2} )$. It is the group $C_{2} \times W(H'_{2} )$
which leaves the unit quaternion 1 invariant. The generators of the
group can be written as $C_{2} \times W(H'_{2} )=\{ [1,1]^{*}
,[\bar{b},b],[e_{3} ,-e_{3} ]^{*} \}.$ The other set $20_{-}$ has a
similar decomposition. Each set of quaternions $10_{+} {\rm \; and\;
}10'_{+}$ can be written as

\begin{equation} \label{GrindEQ__31_}
\begin{array}{lll} 10_{+} &=&\{ \frac{1}{2} (1+e_{1} +e_{2} +e_{3} ),{\rm \; }\frac{1}{2} (1-e_{1} -e_{2} -e_{3} ),{\rm \; \; } \\ && \frac{1}{2} (1-e_{1} -e_{2} +e_{3} ),{\rm \; \; }\frac{1}{2} (1+e_{1} +e_{2} -e_{3} ), \\ &&\frac{1}{2} (1\pm \sigma e_{1} \pm \tau e_{3} ){\rm \; ,}\frac{1}{2} (1+\tau e_{1} -\sigma e_{2} ),{\rm \; \; }\frac{1}{2} (1-\tau e_{1} +\sigma e_{2} )\},  \\\\

10^\prime_{+} &=&\{ \frac{1}{2} (1-e_{1} +e_{2} -e_{3} ),{\rm \; }\frac{1}{2} (1+e_{1} -e_{2} -e_{3} ),{\rm \; \; } \\ && \frac{1}{2} (1-e_{1} +e_{2} +e_{3} ),{\rm \; \; }\frac{1}{2} (1+e_{1} -e_{2} -e_{3} ), \\ && \frac{1}{2} (1\pm \tau e_{2} \pm \sigma e_{3} ){\rm \; ,}\frac{1}{2} (1+\tau e_{1} +\sigma e_{2} ),{\rm \; \; }\frac{1}{2} (1-\tau e_{1} -\sigma e_{2} )\}.  \end{array}
\end{equation}

\begin{figure}[h]
\begin{center}
      \subfigure[]{\includegraphics[height=4cm]{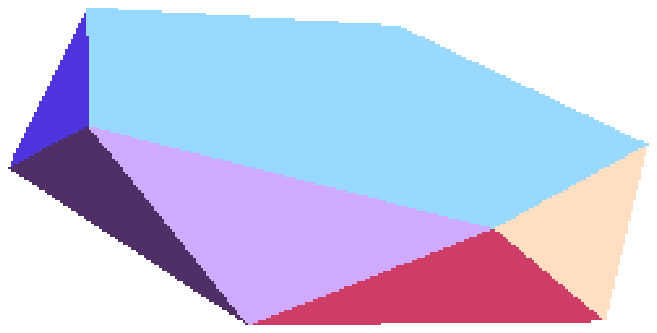}}
      \subfigure[]{\includegraphics[height=4cm]{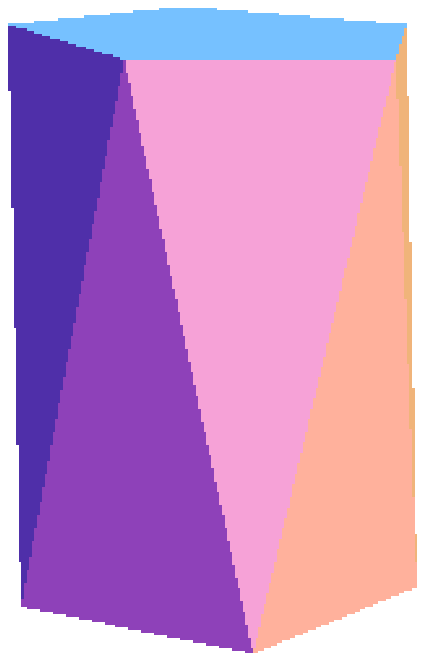}}\\
\end{center}
\caption{Distorted pentagonal antiprisms obtained from a dodecahedron
  (a) the solid represented by the set of vertices of $10_{+}$ and (b)
  the solid represented by the vertices of $10^\prime_{+}$.}
\end{figure}

\noindent The intersection of the hyperplane orthogonal to the
quaternion 1 with the sphere $S^{3}$ at the origin will lead to 20
vertices of the \textit{grand antiprism} given by

\begin{equation} \label{GrindEQ__32_}
\begin{array}{l} {\pm e_{1} ,\pm e_{2} ,\frac{1}{2} (\pm \sigma e_{1} \pm \tau e_{2} \pm e_{3} ),\pm \frac{1}{2} (e_{1} +\sigma e_{2} +\tau e_{3} ),} \\ {\pm \frac{1}{2} (e_{1} +\sigma e_{2} -\tau e_{3} ),\pm \frac{1}{2} (-\tau e_{1} +e_{2} +\sigma e_{3} ),\pm \frac{1}{2} (\tau e_{1} -e_{2} +\sigma e_{3} )}. \end{array}
\end{equation}
\noindent 

\noindent These are the remaining quaternions of the set 30 of Table 1
after removing 10 quaternions $\{ e_{3} b^{m}\}, (m=0,1,...,9)$.

\noindent Under the group $C_{2} \times W(H'_{2} )$ the set of
quaternions in (\ref{GrindEQ__32_}) splits as

\begin{equation} \label{GrindEQ__33_}
\begin{array}{lll}
P_{1} &=&\{ \pm e_{1} ,\pm \frac{1}{2} (e_{1} +\sigma e_{2} +\tau e_{3} ),\pm \frac{1}{2} (e_{1} +\sigma e_{2} -\tau e_{3} ), \\
&& \pm \frac{1}{2} (-\sigma e_{1} +\tau e_{2} \pm e_{3} )\},  \\

P_{2} &=&\{ \pm e_{2} ,\pm \frac{1}{2} (\tau e_{1} -e_{2} +\sigma e_{3} ),\pm \frac{1}{2} (-\tau e_{1} +e_{2} +\sigma e_{3} ), \\
&& \pm \frac{1}{2} (\sigma e_{1} +\tau e_{2} \pm e_{3} )\}.
\end{array}
\end{equation}

\noindent The two sets in~(\ref{GrindEQ__33_}) are pentagonal antiprism
and non-regular stretched antiprism respectively  as plotted in Figure~10.

\begin{figure}[h]
\begin{center}
      \subfigure[]{\includegraphics[height=4cm]{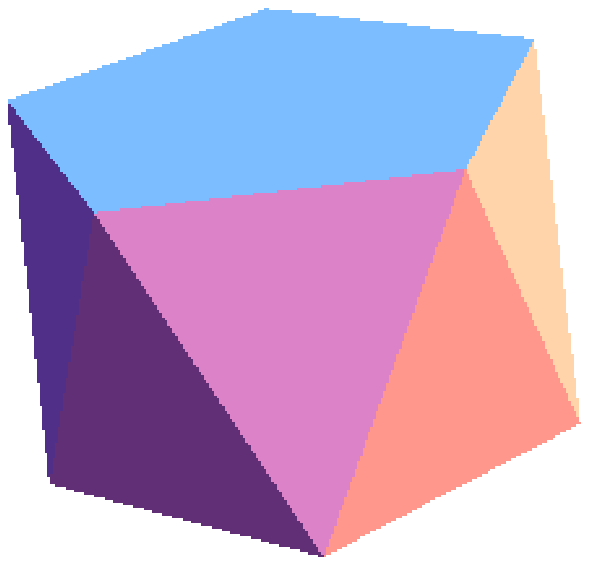}}
      \subfigure[]{\includegraphics[height=4cm]{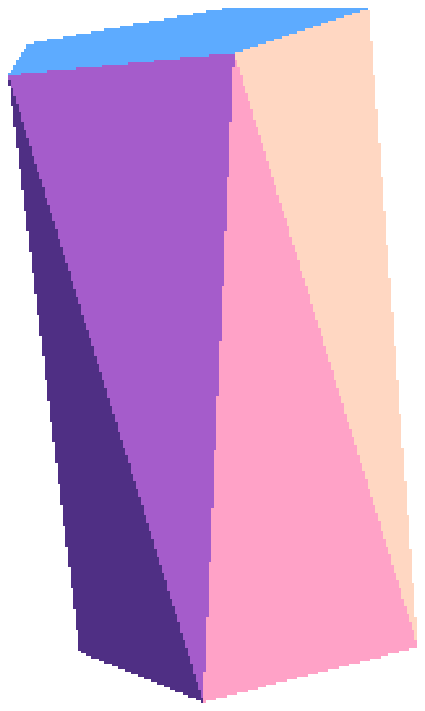}}\\
\end{center}
\caption{(a) Pentagonal antiprism represented by the vertices in (\ref{GrindEQ__33_}) (first line), (b) The solid representing the vertices of (\ref{GrindEQ__33_}) (second line)
}
\end{figure}

\section{Dual of the \textit{grand antiprism}}

\noindent To determine one cell of the \textit{dual grand antiprism
}one should find the centers of the cells which share one vertex of
the \textit{grand antiprism}. Let us choose a particular vertex
represented by the quaternion \textit{c}. We have seen that there are
2 pentagonal antiprisms and 12 tetrahedra sharing the quaternion
\textit{c} as a common vertex. We have already noted that the centers
of two pentagonal antiprisms are $\frac{\tau }{2} 1$ and $\frac{\tau
}{2} b$. The centers of the 12 tetrahedra can be determined by
averaging the corresponding vertices. The centers of the tetrahedra,
up to a scale factor, are given by the following unit quaternions:

\begin{equation} \label{GrindEQ__34_}
\begin{array}{ll} 
c_{1} =\frac{1}{2\sqrt{2}} (\tau +e_{1} +2e_{2} -\sigma e_{3} ), & c_{2} =\frac{1}{2\sqrt{2} } (\tau -\sigma e_{1} +\sqrt{5} e_{2} ), \\ c_{3} =\frac{1}{2\sqrt{2}}  (2-\sigma e_{1} +\tau e_{2} +e_{3} ), & c_{4} =\frac{1}{2\sqrt{2}}(2+\tau e_{1} +e_{2} +\sigma e_{3} ), \\
c_{5} =\frac{1}{2\sqrt{2} } (\sqrt{5} +\tau e_{1} -\sigma e_{2} ), & c_{6} =\frac{1}{2\sqrt{2} } (\sqrt{5} +e_{1} +e_{2} -e_{3} ), \\ c_{7} =\frac{1}{2\sqrt{2}}, (\tau +\tau e_{1} +\tau e_{2} -\sigma ^{2} e_{3} ), & c_{8} =\frac{1}{2\sqrt{2}} (\tau +e_{1} +2e_{2} +\sigma e_{3} ), \\ c_{9} =\frac{1}{2\sqrt{2}} (\tau +\tau e_{1} +\tau e_{2} +\sigma ^{2} e_{3} ), & c_{10} =\frac{1}{2\sqrt{2}}(2+\tau e_{1} +e_{2} -\sigma e_{3} ), \\ c_{11} =\frac{1}{2\sqrt{2} } (\sqrt{5} +e_{1} +e_{2} +e_{3} ), & c_{12} =\frac{1}{2\sqrt{2} } (2-\sigma e_{1} +\tau e_{2} -e_{3} ). 
\end{array} 
\end{equation} 

\noindent One can easily check that any vector $c_{i} -c_{j} , (i\ne
j=1,2,...,12)$ is orthogonal to the vector \textit{c}.  One notes that
the quaternion $1-b$ is also orthogonal to the quaternion~\textit{c}.
However the quaternions $1-c_{i}$ and $b-c_{i}$ are not orthogonal to
the quaternion~\textit{c}.  The relative magnitudes of 1 and
\textit{b} must be fixed to have their tips on the same hyperplane
with the vectors $c_{i}$. One can easily check that the quaternions
$c_{13} =\frac{\tau }{\sqrt{2} } 1$ and $c_{14} =\frac{\tau }{\sqrt{2}
} b=\frac{1}{2\sqrt{2} } (\tau +\sigma e_{1} +e_{2} )$ satisfy the
requirement. These 14 quaternions are in a single hyperplane which is
orthogonal to the vertex \textit{c}. The 14 quaternions $c_{i} ,{\;
  (i=1,2,...,14)}$ represent the vertices of the cell of the
\textit{dual grand antiprism}. It is clear that the vertices of
the\textit{ dual grand antiprism} have two different lengths, 20 of
which are of length $\frac{\tau }{\sqrt{2} }$ and the remaining 300
vertices are of unit length. Therefore, vertices of the\textit{ dual
  grand antiprism} lie on two concentric spheres $S^{3}$ with radii
$\frac{\tau }{\sqrt{2} }$ and~1. The cell can be plotted, similar to
the vertex figure, in the space of intersections of the hyperplane,
orthogonal to the quaternion \textit{c}, with two $S^{3}$ spheres.
This means we need to use the basis introduced
in~(\ref{GrindEQ__27_}).  When 14 vertices $c_{i} ,{(i=1,2,...,14)},$
are expressed in terms of the new basis vectors then all first
components would read $\frac{\tau ^{2} }{2\sqrt{2} } d_{0}$ and the
remaining three components, up to a scale factor $\frac{1}{2\sqrt{2}
}$, can be written as

\begin{equation} \label{GrindEQ__35_}
\begin{array}{ll} c_{1} \approx (-\sigma ,-\sigma ,-\sigma ), & c_{2} \approx (0,1,\sigma ^{2} ), \\ c_{3} \approx (\sigma ^{2} ,0,1), & c_{4} \approx (\sigma ^{2} ,0,-1), \\ c_{5} \approx (-\sigma ,\sigma ,\sigma ), & c_{6} \approx (-\sigma ^{2} ,0,-1) \\ c_{7} \approx (-\sigma ,-\sigma ,\sigma ), & c_{8} \approx (0,1,-\sigma ^{2} ), \\ c_{9} \approx (1,\sigma ^{2} ,0) & c_{10} \approx (1,-\sigma ^{2} ,0), \\ c_{11} \approx (-\sigma ,\sigma ,-\sigma ), & c_{12} \approx (\sigma ,-\sigma ,\sigma ), \\ c_{13} \approx (-1,-\tau ,0), &  c_{14} \approx (-\tau ,0,1). \end{array} 
\end{equation} 

\noindent We obtain the following solid in three dimensions as shown
in Figure~11.

\begin{figure}[h]
\begin{center}
 \includegraphics[height=4cm]{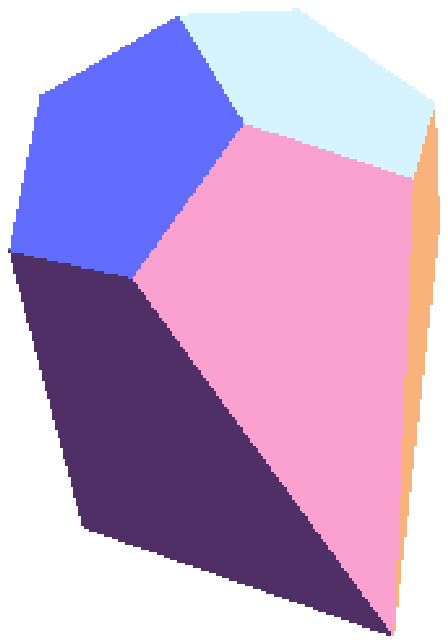}
\end{center}
\caption{One cell of the \textit{dual grand antiprism}}
\end{figure}

\noindent This cell has the same symmetry $C_{2} \times C_{2}$ of the
vertex figure which fixes the quaternion \textit{c}. The following
sets of vertices of the cell are left invariant under the action of
$C_{2} \times C_{2}$;

\noindent 

\begin{equation} \label{GrindEQ__36_} 
(c_{1} ,c_{4} ,c_{8} ,c_{10} ),{\rm \; }(c_{2} ,c_{5} ),{\rm \; }(c_{3} ,c_{6} ,c_{11} ,c_{12} ),{\rm \; \; }(c_{7} ,c_{9} ),{\rm \; }(c_{13} ,c_{14} ).                        
\end{equation} 

\noindent One cell of the \textit{dual grand antiprism }consists of
four regular pentagons, four kites and two isosceles trapezoids where
the edge lengths are shown in Figure~12.

\begin{figure}[h]
\begin{center}
  \includegraphics[height=4cm]{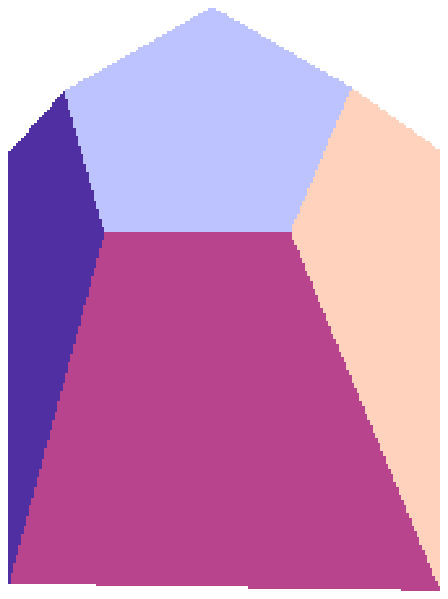}
\end{center}
\caption{Faces of the cell of {\it dual grand antiprism}}
\end{figure}

\noindent To visualize the cell of the \textit{dual grand antisprism
}we note that the following sets of vertices form the faces of the
cell:

\noindent
\begin{equation} \label{GrindEQ__37_}
\begin{array}{l}
{\rm Pentagons:} \\
(c_{1} ,c_{2} ,c_{8} ,c_{7} ,c_{9} ),{\rm \; \; }(c_{8} ,c_{12} ,c_{6} ,c_{4} ,c_{7} ),{\rm \; \; } \\ (c_{7} ,c_{4} ,c_{5} ,c_{10} ,c_{9} ),{\rm \; \; }(c_{9} ,c_{10} ,c_{11} ,c_{3} ,c_{1} ); \\

{\rm Kites:} \\
(c_{14} ,c_{2} ,c_{1} ,c_{3} ),{\rm \; \; (}c_{14} ,c_{2} ,c_{8} ,c_{12} ),{\rm \; \; (}c_{13} ,c_{5} ,c_{4} ,c_{6} ),{\rm \; \; (}c_{13} ,c_{11} ,c_{10} ,c_{5} ); \\

{\rm Isosceles~trapezoids:} \\
(c_{13} ,c_{14} ,c_{12} ,c_{6} ),{\rm \; \; (}c_{13} ,c_{14} ,c_{3} ,c_{11} ).
\end{array}
\end{equation}

\noindent Let us note that the pair of vertices $(c_{13} ,c_{14} )\in
H_{2}$ is left invariant by a group larger than the group $C_{2}
\times C_{2}$. Indeed the largest group fixing the set $(c_{13}
,c_{14} )\in H_{2}$ is the direct product of two groups $C_{2} \times
W(H'_{2} )$ where the first group is generated by the group element
$[\bar{b}^{2} ,-\bar{b}^{2} ]^{*} \in W(H_{2} )$. The generators of
the group $C_{2} \times W(H'_{2} )$ can be chosen as follows

\begin{equation} \label{GrindEQ__38_} 
C_{2} \times W(H'_{2} )=\{ [\bar{b},b],[e_{3} ,-e_{3} ]^{*} ,[\bar{b}^{2} ,-\bar{b}^{2} ]^{*} \} .        
\end{equation} 

\noindent The center of the cell of the {\it dual grand antiprism}
represented by the vertex \textit{c}, as we know, is left invariant by
two generators in~(\ref{GrindEQ__38_}). The generator $[\bar{b},b]$
generates the five vertices $\bar{b}^{m} cb^{m} ,{\; \;
  (m=0,1,2,3,4)}$ of the \textit{grand antiprism} when applied to
\textit{c}. Action of the group element $[\bar{b},b]$ on the other
vertices $c_{i} ,{\; (i=1,2,...,12)}$ will generate other cells joined
to the vertices $(c_{13} ,c_{14} )\in H_{2}$. That is to say we
generate five cells all sharing the vertices $c_{13}$ and $c_{14} $.
For each pair of quaternions $(b^{m} ,b^{m+2} ),{\; (m=0,1,...,9)}$ we
have $5\times 10=50$ cells overall and $(e_{3} b^{m} ,e_{3} b^{m+2}
),{\; (m=0,1,...,9)}$ another 50 cells with a total of 100 cells of
the \textit{dual grand antiprism}. This is of course the number of
vertices of the\textit{ grand antiprism.} As we noted before, the dual
grand antiprism has 20 vertices belonging to the root system $H_{2}
\oplus H'_{2}$. The remaining 300 vertices belong to two different
orbits of $Aut(H_{2} \oplus H'_{2} )$ of sizes 200 and 100.  They can
be computed as follows. In the earlier paper~[10] we determined the
600 vertices of the 120-cell as the set of quaternions $J=\sum
_{i,j=o}^{5}b^{i} T'b^{j}$ where $T'=V_{1} \oplus V_{2} \oplus V_{3}$
represents the vertices of the 24-cell with $V_{1} ,V_{2} ,V_{3}$
given by

\begin{equation} \label{GrindEQ__39_} 
\begin{array}{l} {V_{1} =\{ \frac{1}{\sqrt{2} } (\pm 1\pm e_{1} ),\frac{1}{\sqrt{2} } (\pm e_{2} \pm e_{3} )\},} \\ {V_{2} =\{ \frac{1}{\sqrt{2} } (\pm 1\pm e_{2} ),\frac{1}{\sqrt{2} } (\pm e_{3} \pm e_{1} )\} ,} \\ {V_{3} =\{ \frac{1}{\sqrt{2} } (\pm 1\pm e_{3} ),\frac{1}{\sqrt{2} } (\pm e_{1} \pm e_{2} )\}.} \end{array} 
\end{equation} 

\noindent It is not difficult to see that the 600 vertices of the
120-cell decompose under the group $Aut(H_{2} \oplus H'_{2} )$ as
600=200+200+100+100.  With the choice of $b=\frac{1}{2} (\tau +\sigma
e_{1} +e_{2} )$, we see that the 300 vertices of the \textit{dual
  grand antiprism} are given by the set of quaternions

\begin{equation} \label{GrindEQ__40_} 
\begin{array}{l}
J_{1} =\sum _{i,j=0}^{5}b^{i} V_{1} b^{j}  {\rm ~of~ size~ 200,} \\ \\ 
J'_{3} =\sum _{i,j=0}^{5}b^{i} \frac{1}{\sqrt{2} }  (\pm 1\pm e_{3} )b^{j} {\rm ~of~ size~ 100}.
\end{array}
\end{equation}
\noindent 

\noindent One can check that in~(\ref{GrindEQ__35_}) the first 12
vertices belong to the set of 20 vertices of a dodecahedron and form
four pentagonal faces only instead of 12 pentagonal faces of a
dodecahedron. This is because 8 vertices are missing from the
dodecahedron which is a typical cell of the 120-cell.  For each vertex
of the \textit{grand antiprism} we have 12 vertices forming 4
pentagons which is part of the dual cell, then the number of vertices
of the \textit{dual grand antiprism} belonging to the set J is
$\frac{100\times 12}{4} =300$ as we have explained before which is the
union of the sets $J_{1}$ and $J'_{3} $.

\section{Conclusion}

\noindent We have constructed the \textit{grand antiprism} and its
dual polytope with quaternions as a subset of the vertices of the
600-cell represented by the quaternionic elements of the binary
icosahedral group. It is a 4D semi-regular polytope with 100 vertices
and 320 cells made of 300 tetrahedral and 20 pentagonal antiprisms.
Construction of its vertices and its symmetry group elements in terms
of quaternions is very simple and very elegant. One needs only two
orthogonal quaternions $b{\rm \; and}{\; e_{3}}$ to obtain the group
elements and an additional quaternion \textit{c} for the construction
of the vertices. The originality of the work lies in the fact that
both the vertices of the grand antiprism and its symmetry group are
constructed in terms of quaternions. We also note that the dual
polytope of the grand antiprism has not been constructed elsewhere.

\section*{Acknowledgement}

\noindent We would like to thank Dr. T. Yashiro and Dr. B. Anchouche
for discussions.

\newpage
\begin{appendix}
\section*{Appendix: Decomposition of the semiregular orbits of $W(H_{4} )$ under
the its maximal subgroup $Aut(H_{2} \oplus H'_{2} )\approx \{ W(H_{2}
)\times W(H'_{2} )\} :C_{4}$}

Orbit 1:\\ 600 =
2(100)+
2(200) \\
Orbit 2:\\ 1200 =
3(200)+
2(100)+
400. \\
Orbit 3:\\ 720 =
100+
3(200)+
20. \\
Orbit 4:\\ 120 =
20+
100. \\
Orbit 5:\\ 3600 =
6(400)+
5(200)+
2(100). \\
Orbit 6:\\ 2400 =
6(200)+
3(400). \\
Orbit 7:\\ 3600 =
5(200)+
6(400)+
2(100). \\
Orbit 8:\\ 1440 =
5(200)+
400+
40. \\
Orbit 9:\\ 2400 =
3(400)+
6(200). \\
Orbit 10:\\ 3600 =
6(400)+
5(200)+
2(100). \\
Orbit 11:\\ 7200 =
6(200)+
15(400). \\
Orbit 12:\\ 7200 =
15(400)+
6(200). \\
Orbit 13:\\ 7200 =
15(400)+
6(200). \\
Orbit 14:\\ 7200 =
15(400)+
6(200). \\
Orbit 15:\\ 14400 =
36(400). \\
\end{appendix}

\addcontentsline{toc}{chapter}{Bibliography}

\markright{Bibliography}


\begin{thebibliography}{99}

\bibitem{1} M. Koca, R. Koc, M.Al-Barwani, J.M.Phys. 44 (2003) 03123;
  M. Koca, R. Koc, M. Al-Barwani, J. M. Phys 47 (2006) 043507-1;
  M.Koca, R. Koc, M. Al-Ajmi, J. Phys. A: Math. Gen. 39 (2006) 14047.
\bibitem{2} H. Georgi and S.L. Glashow, Phys Rev. Lett. 32 ( 1974)
  438.
\bibitem{3} T. Pengpan and P.Ramond, Phys.Rep.C\textbf{315} (1999)
  137; P.Ramond, ``Algebraic Dreams'', UFIFT-HET-01-27.
\bibitem{4} H. S. M. Coxeter, Regular Polytopes(3rd Edition), Dover
  Publications, INC. New York, 1973 ; H. S. M. Coxeter, Regular
  Complex Polytopes, Cambridge: Cambridge University Press,
  1973.
\bibitem{5} P. du Val, Homographies, Quaternions and Rotations, Oxford
  University Press,1964;
\bibitem{6} J.H. Conway and D. A. Smith, On Quaternions and Octonions:
  their geometry, arithmetic, and symmetry, A. K. Peters Ltd, Natic,
  Massachusetts.
\bibitem{7} V. Elsver, N. J. A. Sloane, J. Phys. A 20 (1987) 6161; M.
  Koca, J. Phys.: Math.Gen. \textbf{\underbar{A22}} (1989) 1949; ibid
  J.Phys.  Math.Gen.\textbf{ A22} (1989) 4125; R.V. Moody and J.
  Patera, J Phys A: Math Gen. 26 (1993)2829.
\bibitem{8} M. Koca, R. Koc, M. Al-Barwani,J.Phys.A 34 (2001)11201.
\bibitem{9} J. H. Conway and Guy, \textit{Four-Dimensional Archimedean
    Polytopes}, Proceedings of the Colloquium on Convexity at
  Copenhagen, page 38 und 39, 1965.  ;
  http://en.wikipedia.org/wiki/Uniform-polycoron;
\bibitem{10} M. Koca, R. Koc and M. Al- Ajmi, J. Phys. A: Math. Theor.
  40 (2007) 7633.
\bibitem{11} M. Koca, R. Koc, M.Al-Barwani and S. Al-Farsi, Linear
  Algeb. Appl. 412 (2006) 441.
\bibitem{12} M. Koca, M. Al-Ajmi and R. Koc, J. M. Phys., 48 (2007)
  113514.

\end{thebibliography}
\end{document}